\documentclass[aps,prl,twocolumn,letterpaper]{revtex4}

\usepackage{amsmath,amssymb,amsfonts,amsthm}
\usepackage{graphicx}

\begin{document}

\title{BaNiF$_4$: an electric field-switchable weak antiferromagnet}

\date{\today}

\author{Claude Ederer}
\affiliation{Materials Department, University of California, Santa
  Barbara, CA 93106-5050, U.S.A.}
\email{ederer@mrl.ucsb.edu}
\author{Nicola A.~Spaldin}
\affiliation{Materials Department, University of California, Santa
  Barbara, CA 93106-5050, U.S.A.}
 
\begin{abstract}
We show that in the antiferromagnetic ferroelectric BaNiF$_4$ the
Dzyaloshinskii-Moriya interaction leads to a small canting of the
magnetic moments, away from the easy axis, resulting in a noncollinear
magnetic structure. The canting corresponds to an additional ``weak''
antiferromagnetic order parameter whose orientation is determined by
the polar structural distortion and can be reversed by switching the
ferroelectric polarization with an electric field. Our results point
the way to a more general coupling mechanism between structural
distortions and magnetic order parameters in magnetoelectric
multiferroics, which can be exploited in the design of electric
field-switchable magnets.
\end{abstract}


\maketitle

There is great interest in magnetic ferroelectrics, which primarily
stems from the possible cross-correlations between their magnetic and
dielectric properties \cite{Fiebig:2005}. Such \emph{magnetoelectric
coupling} facilitates the manipulation of magnetic properties using an
electric field and vice versa. Various intriguing effects have been
reported so far, ranging from a magnetic field dependence of the
dielectric permittivity in various materials including YMnO$_3$
\cite{Huang_et_al:1997}, to magnetic phase control using an electric
field in HoMnO$_3$ \cite{Lottermoser_et_al:2004}, and the
reorientation of electric polarization by a magnetic field in
TbMnO$_3$ \cite{Kimura_et_al_Nature:2003}.

Of particular appeal, both from the viewpoint of fundamental science
and also with respect to possible applications in digital memory
technologies, is the switching of \emph{magnetic} domains by an
\emph{electric} field and vice versa. Here, the field is used to push
the system into a different realization of the same ground state
phase. The resulting state continues to be stable even when the field
is removed, thus exhibiting the basic requirement for nonvolatile data
storage. Such magnetoelectric domain switching has been demonstrated
for multiferroic Ni$_3$B$_7$O$_{13}$I, where an electric field can be
used to rotate the magnetization by $\pm$90$^\circ$ and a magnetic
field can switch the polarization into any possible domain
configuration \cite{Ascher_et_al:1966}. Nevertheless, despite the
recent revival of interest in magnetoelectric phenomena, this has, to
our knowledge, remained the only report of magnetoelectric domain
switching in a single-phase multiferroic. In many multiferroics
different mechanisms are driving the magnetic and ferroelectric orders
respectively, resulting in only weak coupling between the two order
parameters, and disfavoring the possibility of magnetoelectric domain
switching. However, electric field-induced magnetization switching has
been observed recently in a nano-composite consisting of ferrimagnetic
CoFe$_2$O$_4$ pillars embedded in an antiferromagnetic and
ferroelectric BiFeO$_3$ matrix \cite{Zavaliche_et_al:2005}.

In this letter we introduce a mechanism in which the magnetic order is
induced by the ferroelectricity and is therefore intimately coupled to
the polarization, allowing controlled switching of the magnetic order
parameter by switching the electric polarization using an electric
field. We demonstrate this approach using as an example the magnetic
ferroelectric BaNiF$_4$, which is representative of a whole class of
multiferroic barium fluorides. We show by using both first-principles
calculations and symmetry analysis that the polar distortion in this
material gives rise to an additional ``weak'' antiferromagnetic order
parameter which is reversed when the ferroelectric polarization is
reversed. We suggest the possible experimental observation of this
electric field-switchable weak antiferromagnetism and discuss the
general implications of our results for magnetoelectric switching
phenomena.

\begin{figure}
\centerline{\includegraphics[width=0.85\columnwidth]{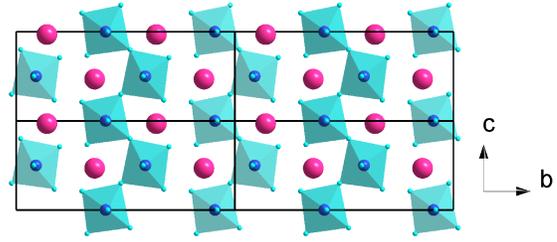}}
\caption{(Color online) Projection of the Ba$M$F$_4$ structure along
  the $a$ axis. The $M$ cations are surrounded by fluorine octahedra,
  which form puckered sheets perpendicular to the $b$ axis, separated
  by similar sheets of Ba cations. Note that adjacent sheets are
  shifted relative to each other by half a lattice constant along the
  $a$ direction.}
\label{fig:struct}
\end{figure}

\begin{figure}
\centerline{\includegraphics[width=0.8\columnwidth]{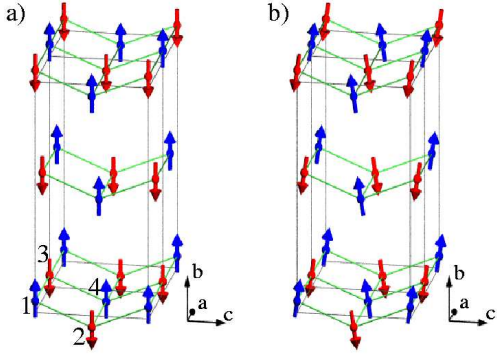}}
\caption{(Color online) Magnetic structures (not to scale) of
  BaNiF$_4$ derived from the experimental observations (a) and from
  our calculation including spin-orbit coupling (b). Gray lines
  outline the conventional orthorhombic unit cell, green lines show
  the puckered sheets perpendicular to the $b$ direction, numbers in
  (a) indicate the 4 magnetic ions in the unit cell.}
\label{fig:mag}
\end{figure}

BaNiF$_4$ is representative of the isostructural family of barium
fluorides with the chemical formula Ba$M$F$_4$, where $M$ can be Mn,
Fe, Co, Ni, Zn, or Mg. These compounds were first synthesized in 1968
\cite{Eibschuetz/Guggenheim:1968, Schnering/Bleckmann:1968}, and
crystallize in a base-centered orthorhombic structure with space group
$Cmc2_1$ \cite{Schnering/Bleckmann:1968}
\footnote{We use the notation of Ref.~\cite{Bradley/Cracknell:Book}
for crystallographic and magnetic space groups.}, which is shown in
Fig.~\ref{fig:struct}. BaMnF$_4$ undergoes an additional structural
phase transition at around 250\,K (see
Ref.~\onlinecite{Scott:1979}). Ferroelectric switching has been
reported for $M$=Co, Ni, Zn, and Mg but not for $M$=Mn and Fe
\cite{Eibschuetz_et_al:1969}. The systems with $M$=Mn, Fe, Co and Ni
order antiferromagnetically at N{\'e}el temperatures around
20-70\,K. The magnetic structure derived experimentally for the
systems with $M$=Mn, Fe, Ni is shown in Fig.~\ref{fig:mag}a
\cite{Cox_et_al:1970}. The magnetic unit cell is doubled compared to
the chemical unit cell, and contains four magnetic $M$ cations, which
are arranged in sheets perpendicular to the $b$ axis. Within each
sheet the cations form a puckered rectangular grid, with the magnetic
moments of neighboring cations oriented antiferromagnetically, and all
moments aligned parallel to the $b$ axis ($c$ axis for $M$=Co
\cite{Eibschuetz_et_al:1972}). The coupling between different sheets
is weak, leading to low magnetic ordering temperatures and pronounced
two-dimensional behavior \cite{Eibschuetz_et_al:1972}. The
corresponding magnetic space group is $P_a2_1$ ($P_a2'_1$ for
BaCoF$_4$).

In this letter we focus on the intriguing magnetoelectric and
magnetostructural properties of BaNiF$_4$. A detailed comparative
study of the structural, electronic, and magnetic properties of the
whole series of multiferroic compounds will be reported elsewhere
\cite{Ederer/Spaldin:unpublished}. All calculations presented in this
work are performed using the {\sc Vienna Ab-Initio Simulation Package}
(VASP) \cite{Kresse/Furthmueller_PRB:1996} employing the
projector-augmented wave method
\cite{Bloechl:1994,Kresse/Joubert:1999}. To account for the strong
Coulomb interaction between the localized $d$ electrons of the
transition metal cations we use the LSDA+$U$ method
\cite{Anisimov/Aryatesiawan/Liechtenstein:1997}. For the Hubbard $U$
and intra-atomic exchange parameter $J$ we use typical values of
$U$=4\,eV and $J$=1\,eV, and we check the sensitivity of our results
with respect to variations of both $U$ and $J$. We use a plane-wave
energy cutoff of 450\,eV and a $\Gamma$-centered 2$\times$4$\times$2
k-point mesh (divisions with respect to the monoclinic basis vectors
of the magnetic unit cell), which is sufficient to obtain converged
results for all quantities under consideration.

First we perform a full structural optimization of BaNiF$_4$ within
the experimentally observed $Cmc2_1$ symmetry; the structural
parameters we obtain agree well with available experimental data
\cite{Ederer/Spaldin:unpublished}. To confirm the experimentally
reported magnetic structure we then calculate the energy differences
for different relative orientations of the four magnetic sublattices
corresponding to the four magnetic ions in the unit cell and extract
the Heisenberg nearest neighbor coupling constants. In these
calculations the spin-orbit coupling is neglected. We obtain strong
antiferromagnetic nearest neighbor coupling within the buckled planes,
$J_a$=4.4\,meV, $J_c$=3.3\,meV \footnote{We use the form $E_{ij} =
-2J_{ij} s_i \cdot s_j$ for the exchange interaction; $J_a = J_{13}$,
$J_c = J_{12}$, referring to the indices in Fig.~\ref{fig:mag}a.},
consistent with the experimentally observed structure. However, when
spin-orbit coupling is included in the calculation, we observe that
the collinear spin-configuration shown in Fig.~\ref{fig:mag}a, with
all spins aligned along the $b$ direction, is unstable, and that
instead the magnetic moments assume a noncollinear configuration where
all spins are slightly tilted towards the $\pm c$ direction as shown
in Fig.~\ref{fig:mag}b. The tilting angle is about 3\,$^\circ$.

\begin{figure}
\centerline{\includegraphics[width=0.5\columnwidth]{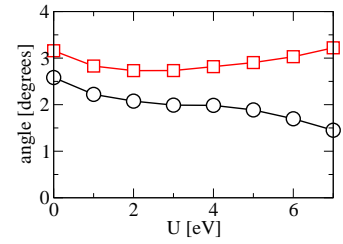}}
\caption{(Color online) Dependence of canting angle on the LSDA+$U$
parameters. Black circles correspond to $J$=0 and red squares to
$J$=1\,eV.}
\label{fig:angle}
\end{figure}

Figure~\ref{fig:angle} shows the $U$ and $J$ dependence of the canting
angle between the magnetic moments of the Ni cations and the $b$
direction. In order to obtain local magnetic moments we integrate the
spin-density within spheres of radius 1.2\,\AA\ centered at the Ni
sites. Our test calculations show that for this radius the integrated
spin-density adopts a saturated value of about 1.7\,$\mu_\text{B}$. It
can be seen that within the physically reasonable range of $U$=2-6\,eV
there is only a moderate $U$ dependence of the canting angle, whereas
it is slightly suppressed by decreasing $J$. This $J$ dependence is a
consequence of the spin-non-diagonal elements introduced into the
effective potential by the LSDA+$U$ energy in its most general form
\cite{Solovyev/Liechtenstein/Terakura:1998}, which is rotationally
invariant with respect to both orbital and spin quantum numbers. It
follows from Fig.~\ref{fig:angle} that within the physically
reasonable parameter range of $U$=2-6\,eV and $J$=0-1\,eV the canting
angle is 2-3\,$^\circ$.

The small tilting of the magnetic ions can be explained by the
antisymmetric exchange or Dzyaloshinskii-Moriya (DM) interaction
\cite{Moriya:1960}, $E^\text{DM}_{ij} = d_{ij} \cdot ( s_i \times s_j
)$, where $s_i$ is the spin of ion $i$ and $d_{ij}$ is the coupling
vector corresponding to the antisymmetric exchange interaction between
ions $i$ and $j$. This interaction occurs only for certain low
symmetries and can give rise to ``weak ferromagnetism'' in otherwise
antiferromagnetic materials, where a canting of the two (or more)
sublattice magnetizations away from the ideal collinear
antiferromagnetic orientation gives rise to a small net magnetization
\cite{Moriya:1960,Ederer/Spaldin:2005}. In the case of BaNiF$_4$ the
magnetic space group does not allow the occurrence of weak
ferromagnetism but nevertheless there is a nonzero DM interaction
$d_c$ between magnetic nearest neighbors along the $c$ direction
\footnote{We take $d_c$ to be the DM vector $d_{12}=-d_{21}$,
referring to the indices shown in Fig.~\ref{fig:mag}a.}, whereas the
DM interaction between neighboring cations along the $a$ direction
vanishes by symmetry.

The canting due to the DM interaction leads to a ``weak''
antiferromagnetic order parameter $L_c = s_1 + s_2 - s_3 - s_4$, in
addition to the experimentally observed (primary) antiferromagnetic
order parameter $L_{ab} = s_1 - s_2 - s_3 + s_4$. On a macroscopic
level the DM interaction leads to a coupling between $L_{ab}$ and
$L_c$ of the form:
\begin{equation}
\label{eq:DM}
E^\text{DM}_\text{macro} = D \cdot ( L_{ab} \times L_c ) \quad ,
\end{equation}
where $D=d_c/2$. If we neglect the magnetic single-ion anisotropy
contribution to the total energy, the canting angle is given by
$\alpha \approx D/4J_c$, from which we obtain a value for $D$ of about
0.7\,meV.

Note that a term of the form (\ref{eq:DM}) can also be caused by
different orientations of the easy axes for the magnetic moments on
different sites. This mechanism is also allowed within the
crystallographic and magnetic symmetry of BaNiF$_4$, in addition to
the DM interaction described above. Since the following analysis of
the magnetoelectric domain switching is independent of the actual
microscopic mechanism leading to the term (\ref{eq:DM}), we do not
separate these two effects further.

In order to analyze the possibility of magnetoelectric domain
switching in BaNiF$_4$ we need to know the symmetry of the
corresponding ``prototype'' phase
\cite{Aizu:1970,Ederer/Spaldin:2005}. Although the paraelectric states
of the Ba$M$F$_4$ systems are not accessible experimentally, since all
crystals melt before undergoing a ferroelectric phase transition, a
nonpolar reference structure has been suggested and used to discuss
the ferroelectric switching properties of these systems
\cite{Keve/Abrahams/Bernstein:1969}. This structure has the nonpolar
space group $Cmcm$ and can be obtained from the ground state $Cmc2_1$
structure by enforcing a mirror symmetry perpendicular to the $c$
axis. Within this symmetry no canting of the magnetic moments is
allowed and the resulting magnetic order corresponds to the collinear
spin arrangement shown in Fig.~\ref{fig:mag}a with the magnetic space
group $P_a2_1/m'$. The absence of canting in the centrosymmetric
reference structure indicates an intimate connection between the
structural distortion and the weak antiferromagnetic order parameter
$L_c$.

\begin{figure}
\centerline{\includegraphics[width=0.82\columnwidth]{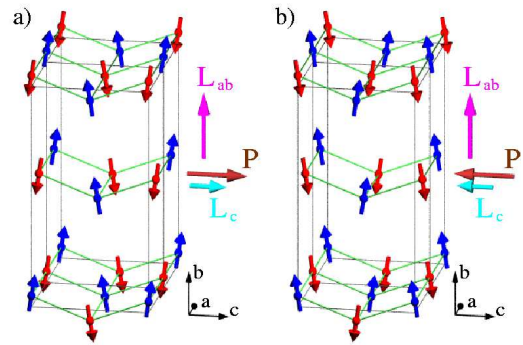}}
\caption{(Color online) Magnetoelectric switching in BaNiF$_4$. The
  weak antiferromagnetic order parameter $L_c$ is coupled to the
  polarization $P$. Reversal of the polarization from (a) to (b) leads
  to a reversal of the canting of the magnetic moments and thus to a
  reversal of $L_c$.}
\label{fig:switch}
\end{figure}

In BaNiF$_4$ only two different structural domains are possible,
corresponding to the two opposite orientations of the ferroelectric
polarization along $\pm c$. Indeed, if we invert the polar mode in our
calculation, we observe that for fixed orientation of $L_{ab}$ the
orientation of the weak antiferromagnetic order parameter is
determined by the orientation of the polarization $P$ (see
Fig.~\ref{fig:switch}). This means that a reversal of $P$ by an
electric field is accompanied by a reversal of $L_c$, if during the
polarization switching the primary antiferromagnetic order parameter
$L_{ab}$ is preserved. Energetically, a reversal of $L_c$ is much more
favorable than reversal of $L_{ab}$, since the latter requires the
magnetic moments to rotate by $\sim$ 180\,$^\circ$ through the hard
magnetic axis, whereas the reversal of $L_c$ requires only a slight
reorientation of the magnetic moments through the easy axis. We
therefore propose that the ferroelectric switching in BaNiF$_4$ will
be accompanied by a reversal of the weak antiferromagnetic order
parameter.

A canted antiferromagnetic structure caused by the antisymmetric
exchange interaction has also been proposed in the seminal paper by
Moriya \cite{Moriya:1960} for CuCl$_2\cdot$2H$_2$O, and has
subsequently been confirmed by several experimental methods including
neutron scattering \cite{Umebayashi_et_al:1967}, antiferromagnetic
resonance measurements \cite{Eremenko_et_al:1983}, and NMR
\cite{Kubo/Yamahaku:1985}. This shows that a ``hidden'' order
parameter such as $L_c$ can be detected experimentally. However, in
order to observe the predicted reversal of $L_c$ by an electric field,
an experimental technique is required that allows the distinction
between different antiferromagnetic 180\,$^\circ$ domains. One such
technique, which in addition allows imaging of the antiferromagnetic
domain topology, is second-harmonic generation (SHG)
\cite{Fiebig/Pavlov/Pisarev:2005}, which, in fact, has already
successfully detected weak antiferromagnetism in YMnO$_3$
\cite{Degenhardt_et_al:2001}. We therefore hope that our work will
stimulate experimental efforts to measure the predicted
magnetoelectric domain switching in BaNiF$_4$ and related materials.

Magnetic order that is induced by ferroelectricity via the
magnetoelectric effect has been suggested for the related system
BaMnF$_4$, based on macroscopic symmetry considerations
\cite{Fox/Scott:1977}. In BaMnF$_4$ an additional structural phase
transition lowers the symmetry so as to allow the linear
magnetoelectric effect, which is symmetry-forbidden in
BaNiF$_4$. Nevertheless, the microscopic mechanism governing
magnetoelectric behavior could be the same in both systems (although a
rigorous microscopic study is required to show this
unambiguously). The weak antiferromagnetism in BaNiF$_4$ would then be
classified as ``anti-magnetoelectric effect'' \cite{Fiebig:2005},
mediated by the DM interaction. A related scenario has also been
reported recently for multiferroic BiFeO$_3$, where the DM interaction
leads to weak ferromagnetism \cite{Ederer/Spaldin:2005}. Although the
linear magnetoelectric effect is symmetry-allowed in this system
\footnote{If the long-wavelength spin rotation is suppressed, see
Ref~\onlinecite{Ederer/Spaldin:2005}.}, first principles calculations
show that the weak ferromagnetic moment in BiFeO$_3$ is in fact not
induced by the ferroelectricity, but is coupled to a non-polar
structural mode \cite{Ederer/Spaldin:2005}, which exemplifies that
symmetry considerations alone cannot provide full insight into the
exact coupling mechanism of magnetoelectric systems.

Since the DM interaction can only be nonzero if the midpoint between
two magnetic ions is not an inversion center, the occurrence of
``weak'' magnetic order due to spin-canting should be a rather common
phenomenon in magnetic ferroelectrics. The resulting order parameter
can in general be ferro- or antiferromagnetic and can be reversed by
reversing the structural mode that is responsible for the nonzero DM
interaction. If this mode is polar, it can be reversed by an electric
field, thus leading to an electric field-switchable magnetic order
parameter. Our results for BaNiF$_4$ and our earlier study of
BiFeO$_3$ show that the amplitude of the corresponding order
parameters are typically of the order of 0.1\,$\mu_\text{B}$ per
magnetic ion, which is about 10\,\% of a typical ``strong'' magnetic
order parameter ($\sim$ 1\,$\mu_\text{B}$ per magnetic ion), which is
easy to measure with modern experimental techniques.

Finally, we compare this with a somewhat converse effect, which is
currently attracting a lot of attention, namely the appearance of a
small electric polarization in materials where the magnetic order
breaks the inversion symmetry of the system. The resulting
polarization is then intimately coupled to the magnetic properties
\cite{Kimura_et_al_Nature:2003,Lawes_et_al:2005}. Although the
underlying microscopic mechanism is not yet understood, the important
role that symmetry plays for this effect is similar to the case of the
weak magnetic order discussed in this letter. The magnitude of the
resulting electric polarization is around 0.01\,$\mu$C/cm$^2$
\cite{Kimura_et_al_Nature:2003,Lawes_et_al:2005}, which is two to
three orders of magnitudes smaller than the spontaneous polarization
in typical ``strong'' ferroelectrics ($\sim$1-10 $\mu$C/cm$^2$). From
this it becomes clear that the ``weak'' magnetic order that is caused
by structural distortions is in comparison a rather large effect and
therefore provides a very promising route for achieving practical
magnetoelectric coupling effects. In addition, since the weak magnetic
order parameter and the ferroelectric polarization are coupled
\emph{by symmetry}, it is not necessary that the primary magnetic and
ferroelectric order temperatures are close together. The weak magnetic
order appears at the N{\'e}el temperature but is in fact caused by the
structural distortions, which can ``freeze in'' at much higher
temperatures.

In summary, we have shown that the multiferroic system BaNiF$_4$
exhibits an additional ``weak'' antiferromagnetic order parameter that
is coupled to the polar distortion, and can be reversed by reversing
the ferroelectric polarization using an electric field. We propose
this mechanism as a promising route for achieving magnetoelectric
switching behavior in ferroelectric magnets.

This work was supported by the NSF's \emph{Chemical Bonding Centers}
program, Grant No. CHE-0434567 and used central facilities provided by
the NSF-MRSEC Award No. DMR05-20415. The authors thank J.~F.~Scott for
valuable discussions.

\bibliography{../../literature.bib}

\begin{thebibliography}{30}
\expandafter\ifx\csname natexlab\endcsname\relax\def\natexlab#1{#1}\fi
\expandafter\ifx\csname bibnamefont\endcsname\relax
  \def\bibnamefont#1{#1}\fi
\expandafter\ifx\csname bibfnamefont\endcsname\relax
  \def\bibfnamefont#1{#1}\fi
\expandafter\ifx\csname citenamefont\endcsname\relax
  \def\citenamefont#1{#1}\fi
\expandafter\ifx\csname url\endcsname\relax
  \def\url#1{\texttt{#1}}\fi
\expandafter\ifx\csname urlprefix\endcsname\relax\def\urlprefix{URL }\fi
\providecommand{\bibinfo}[2]{#2}
\providecommand{\eprint}[2][]{\url{#2}}

\bibitem[{\citenamefont{Fiebig}(2005)}]{Fiebig:2005}
\bibinfo{author}{\bibfnamefont{M.}~\bibnamefont{Fiebig}},
  \bibinfo{journal}{J.~Phys. D: Appl. Phys.} \textbf{\bibinfo{volume}{38}},
  \bibinfo{pages}{R123} (\bibinfo{year}{2005}).

\bibitem[{\citenamefont{Huang et~al.}(1997)\citenamefont{Huang, Cao, Sun, Xue,
  and Chu}}]{Huang_et_al:1997}
\bibinfo{author}{\bibfnamefont{Z.~J.} \bibnamefont{Huang}},
  \bibinfo{author}{\bibfnamefont{Y.}~\bibnamefont{Cao}},
  \bibinfo{author}{\bibfnamefont{Y.~Y.} \bibnamefont{Sun}},
  \bibinfo{author}{\bibfnamefont{Y.~Y.} \bibnamefont{Xue}}, \bibnamefont{and}
  \bibinfo{author}{\bibfnamefont{C.~W.} \bibnamefont{Chu}},
  \bibinfo{journal}{Phys. Rev. B} \textbf{\bibinfo{volume}{56}},
  \bibinfo{pages}{2623} (\bibinfo{year}{1997}).

\bibitem[{\citenamefont{Lottermoser et~al.}(2004)\citenamefont{Lottermoser,
  Lonkai, Amann, Hohlwein, Ihringer, and Fiebig}}]{Lottermoser_et_al:2004}
\bibinfo{author}{\bibfnamefont{T.}~\bibnamefont{Lottermoser}}
\bibnamefont{et~al.}
  \bibinfo{journal}{Nature (London)} \textbf{\bibinfo{volume}{430}},
  \bibinfo{pages}{541} (\bibinfo{year}{2004}).

\bibitem[{\citenamefont{Kimura et~al.}(2003)\citenamefont{Kimura, Goto,
  Shintani, Ishizaka, Arima, and Tokura}}]{Kimura_et_al_Nature:2003}
\bibinfo{author}{\bibfnamefont{T.}~\bibnamefont{Kimura}}
\bibnamefont{et~al.}
  \bibinfo{journal}{Nature (London)} \textbf{\bibinfo{volume}{426}},
  \bibinfo{pages}{55} (\bibinfo{year}{2003}).

\bibitem[{\citenamefont{Ascher et~al.}(1966)\citenamefont{Ascher, Rieder,
  Schmid, and St{\"o}ssel}}]{Ascher_et_al:1966}
\bibinfo{author}{\bibfnamefont{E.}~\bibnamefont{Ascher}},
  \bibinfo{author}{\bibfnamefont{H.}~\bibnamefont{Rieder}},
  \bibinfo{author}{\bibfnamefont{H.}~\bibnamefont{Schmid}}, \bibnamefont{and}
  \bibinfo{author}{\bibfnamefont{H.}~\bibnamefont{St{\"o}ssel}},
  \bibinfo{journal}{J.~Appl. Phys.} \textbf{\bibinfo{volume}{37}},
  \bibinfo{pages}{1404} (\bibinfo{year}{1966}).

\bibitem[{\citenamefont{Zavaliche et~al.}(2005)\citenamefont{Zavaliche, Zheng,
  Mohaddes-Ardabili, Yang, Zhan, Shafer, Reilly, Chopdekar, Jia, Wright
  et~al.}}]{Zavaliche_et_al:2005}
\bibinfo{author}{\bibfnamefont{F.}~\bibnamefont{Zavaliche}}
  \bibnamefont{et~al.}, \bibinfo{journal}{Nano Letters}
  \textbf{\bibinfo{volume}{5}}, \bibinfo{pages}{1793} (\bibinfo{year}{2005}).

\bibitem[{\citenamefont{Eibsch{\"u}tz and
  Guggenheim}(1968)}]{Eibschuetz/Guggenheim:1968}
\bibinfo{author}{\bibfnamefont{M.}~\bibnamefont{Eibsch{\"u}tz}}
  \bibnamefont{and} \bibinfo{author}{\bibfnamefont{H.~J.}
  \bibnamefont{Guggenheim}}, \bibinfo{journal}{Solid State Commun.}
  \textbf{\bibinfo{volume}{6}}, \bibinfo{pages}{737} (\bibinfo{year}{1968}).

\bibitem[{\citenamefont{v.~Schnering and
  Bleckmann}(1968)}]{Schnering/Bleckmann:1968}
\bibinfo{author}{\bibfnamefont{H.~G.} \bibnamefont{v.~Schnering}}
  \bibnamefont{and}
  \bibinfo{author}{\bibfnamefont{P.}~\bibnamefont{Bleckmann}},
  \bibinfo{journal}{Naturwissenschaften} \textbf{\bibinfo{volume}{55}},
  \bibinfo{pages}{342} (\bibinfo{year}{1968}).

\bibitem[{\citenamefont{Scott}(1979)}]{Scott:1979}
\bibinfo{author}{\bibfnamefont{J.~F.} \bibnamefont{Scott}},
  \bibinfo{journal}{Rep. Prog. Phys.} \textbf{\bibinfo{volume}{12}},
  \bibinfo{pages}{1055} (\bibinfo{year}{1979}).

\bibitem[{\citenamefont{Eibsch{\"u}tz et~al.}(1969)\citenamefont{Eibsch{\"u}tz,
  Guggenheim, Wemple, Camlibel, and {DiDomenico Jr.}}}]{Eibschuetz_et_al:1969}
\bibinfo{author}{\bibfnamefont{M.}~\bibnamefont{Eibsch{\"u}tz}},
  \bibinfo{author}{\bibfnamefont{H.~J.} \bibnamefont{Guggenheim}},
  \bibinfo{author}{\bibfnamefont{S.~H.} \bibnamefont{Wemple}},
  \bibinfo{author}{\bibfnamefont{I.}~\bibnamefont{Camlibel}}, \bibnamefont{and}
  \bibinfo{author}{\bibfnamefont{M.}~\bibnamefont{{DiDomenico Jr.}}},
  \bibinfo{journal}{Phys. Lett.} \textbf{\bibinfo{volume}{29A}},
  \bibinfo{pages}{409} (\bibinfo{year}{1969}).

\bibitem[{\citenamefont{Cox et~al.}(1970)\citenamefont{Cox, Eibsch{\"u}tz,
  Guggenheim, and Holmes}}]{Cox_et_al:1970}
\bibinfo{author}{\bibfnamefont{D.~E.} \bibnamefont{Cox}},
  \bibinfo{author}{\bibfnamefont{M.}~\bibnamefont{Eibsch{\"u}tz}},
  \bibinfo{author}{\bibfnamefont{H.~J.} \bibnamefont{Guggenheim}},
  \bibnamefont{and} \bibinfo{author}{\bibfnamefont{L.}~\bibnamefont{Holmes}},
  \bibinfo{journal}{J.~Appl. Phys.} \textbf{\bibinfo{volume}{41}},
  \bibinfo{pages}{943} (\bibinfo{year}{1970}).

\bibitem[{\citenamefont{Eibsch{\"u}tz et~al.}(1972)\citenamefont{Eibsch{\"u}tz,
  Holmes, Guggenheim, and Cox}}]{Eibschuetz_et_al:1972}
\bibinfo{author}{\bibfnamefont{M.}~\bibnamefont{Eibsch{\"u}tz}},
  \bibinfo{author}{\bibfnamefont{L.}~\bibnamefont{Holmes}},
  \bibinfo{author}{\bibfnamefont{H.~J.} \bibnamefont{Guggenheim}},
  \bibnamefont{and} \bibinfo{author}{\bibfnamefont{D.~E.} \bibnamefont{Cox}},
  \bibinfo{journal}{Phys. Rev. B} \textbf{\bibinfo{volume}{6}},
  \bibinfo{pages}{2677} (\bibinfo{year}{1972}).

\bibitem[{\citenamefont{Ederer and Spaldin}()}]{Ederer/Spaldin:unpublished}
\bibinfo{author}{\bibfnamefont{C.}~\bibnamefont{Ederer}} \bibnamefont{and}
  \bibinfo{author}{\bibfnamefont{N.~A.} \bibnamefont{Spaldin}},
  \bibinfo{note}{in preparation}.

\bibitem[{\citenamefont{Kresse and
  Furthm{\"u}ller}(1996)}]{Kresse/Furthmueller_PRB:1996}
\bibinfo{author}{\bibfnamefont{G.}~\bibnamefont{Kresse}} \bibnamefont{and}
  \bibinfo{author}{\bibfnamefont{J.}~\bibnamefont{Furthm{\"u}ller}},
  \bibinfo{journal}{Phys. Rev. B} \textbf{\bibinfo{volume}{54}},
  \bibinfo{pages}{11169} (\bibinfo{year}{1996}).

\bibitem[{\citenamefont{Bl{\"o}chl}(1994)}]{Bloechl:1994}
\bibinfo{author}{\bibfnamefont{P.~E.} \bibnamefont{Bl{\"o}chl}},
  \bibinfo{journal}{Phys. Rev. B} \textbf{\bibinfo{volume}{50}},
  \bibinfo{pages}{17953} (\bibinfo{year}{1994}).

\bibitem[{\citenamefont{Kresse and Joubert}(1999)}]{Kresse/Joubert:1999}
\bibinfo{author}{\bibfnamefont{G.}~\bibnamefont{Kresse}} \bibnamefont{and}
  \bibinfo{author}{\bibfnamefont{D.}~\bibnamefont{Joubert}},
  \bibinfo{journal}{Phys. Rev. B} \textbf{\bibinfo{volume}{59}},
  \bibinfo{pages}{1758} (\bibinfo{year}{1999}).

\bibitem[{\citenamefont{Anisimov et~al.}(1997)\citenamefont{Anisimov,
  Aryasetiawan, and Liechtenstein}}]{Anisimov/Aryatesiawan/Liechtenstein:1997}
\bibinfo{author}{\bibfnamefont{V.~I.} \bibnamefont{Anisimov}},
  \bibinfo{author}{\bibfnamefont{F.}~\bibnamefont{Aryasetiawan}},
  \bibnamefont{and} \bibinfo{author}{\bibfnamefont{A.~I.}
  \bibnamefont{Liechtenstein}}, \bibinfo{journal}{J.~Phys.: Condens. Matter}
  \textbf{\bibinfo{volume}{9}}, \bibinfo{pages}{767} (\bibinfo{year}{1997}).

\bibitem[{\citenamefont{Solovyev et~al.}(1998)\citenamefont{Solovyev,
  Liechtenstein, and Terakura}}]{Solovyev/Liechtenstein/Terakura:1998}
\bibinfo{author}{\bibfnamefont{I.~V.} \bibnamefont{Solovyev}},
  \bibinfo{author}{\bibfnamefont{A.~I.} \bibnamefont{Liechtenstein}},
  \bibnamefont{and} \bibinfo{author}{\bibfnamefont{K.}~\bibnamefont{Terakura}},
  \bibinfo{journal}{Phys. Rev. Lett.} \textbf{\bibinfo{volume}{80}},
  \bibinfo{pages}{5758} (\bibinfo{year}{1998}).

\bibitem[{\citenamefont{Moriya}(1960)}]{Moriya:1960}
\bibinfo{author}{\bibfnamefont{T.}~\bibnamefont{Moriya}},
  \bibinfo{journal}{Phys. Rev.} \textbf{\bibinfo{volume}{120}},
  \bibinfo{pages}{91} (\bibinfo{year}{1960}).

\bibitem[{\citenamefont{Ederer and Spaldin}(2005)}]{Ederer/Spaldin:2005}
\bibinfo{author}{\bibfnamefont{C.}~\bibnamefont{Ederer}} \bibnamefont{and}
  \bibinfo{author}{\bibfnamefont{N.~A.} \bibnamefont{Spaldin}},
  \bibinfo{journal}{Phys. Rev. B} \textbf{\bibinfo{volume}{71}},
  \bibinfo{pages}{060401} (\bibinfo{year}{2005}).

\bibitem[{\citenamefont{Aizu}(1970)}]{Aizu:1970}
\bibinfo{author}{\bibfnamefont{K.}~\bibnamefont{Aizu}}, \bibinfo{journal}{Phys.
  Rev. B} \textbf{\bibinfo{volume}{2}}, \bibinfo{pages}{754}
  (\bibinfo{year}{1970}).

\bibitem[{\citenamefont{Keve et~al.}(1969)\citenamefont{Keve, Abrahams, and
  Bernstein}}]{Keve/Abrahams/Bernstein:1969}
\bibinfo{author}{\bibfnamefont{E.~T.} \bibnamefont{Keve}},
  \bibinfo{author}{\bibfnamefont{S.~C.} \bibnamefont{Abrahams}},
  \bibnamefont{and} \bibinfo{author}{\bibfnamefont{J.~L.}
  \bibnamefont{Bernstein}}, \bibinfo{journal}{J.~Chem. Phys.}
  \textbf{\bibinfo{volume}{51}}, \bibinfo{pages}{4928} (\bibinfo{year}{1969}).

\bibitem[{\citenamefont{Umebayashi et~al.}(1967)\citenamefont{Umebayashi,
  Shirane, Frazer, and Cox}}]{Umebayashi_et_al:1967}
\bibinfo{author}{\bibfnamefont{H.}~\bibnamefont{Umebayashi}},
  \bibinfo{author}{\bibfnamefont{G.}~\bibnamefont{Shirane}},
  \bibinfo{author}{\bibfnamefont{B.~C.} \bibnamefont{Frazer}},
  \bibnamefont{and} \bibinfo{author}{\bibfnamefont{D.~E.} \bibnamefont{Cox}},
  \bibinfo{journal}{J.~Appl. Phys.} \textbf{\bibinfo{volume}{38}},
  \bibinfo{pages}{1461} (\bibinfo{year}{1967}).

\bibitem[{\citenamefont{Eremenko et~al.}(1983)\citenamefont{Eremenko, Naumenko,
  Pashkevich, and Pishko}}]{Eremenko_et_al:1983}
\bibinfo{author}{\bibfnamefont{V.~V.} \bibnamefont{Eremenko}},
  \bibinfo{author}{\bibfnamefont{V.~M.} \bibnamefont{Naumenko}},
  \bibinfo{author}{\bibfnamefont{Y.~G.} \bibnamefont{Pashkevich}},
  \bibnamefont{and} \bibinfo{author}{\bibfnamefont{V.~V.}
  \bibnamefont{Pishko}}, \bibinfo{journal}{JETP Letters}
  \textbf{\bibinfo{volume}{38}}, \bibinfo{pages}{112} (\bibinfo{year}{1983}).

\bibitem[{\citenamefont{Kubo and Yamahaku}(1985)}]{Kubo/Yamahaku:1985}
\bibinfo{author}{\bibfnamefont{T.}~\bibnamefont{Kubo}} \bibnamefont{and}
  \bibinfo{author}{\bibfnamefont{H.}~\bibnamefont{Yamahaku}},
  \bibinfo{journal}{Phys. Rev. B} \textbf{\bibinfo{volume}{32}},
  \bibinfo{pages}{1831} (\bibinfo{year}{1985}).

\bibitem[{\citenamefont{Fiebig et~al.}(2005)\citenamefont{Fiebig, Pavlov, and
  Pisarev}}]{Fiebig/Pavlov/Pisarev:2005}
\bibinfo{author}{\bibfnamefont{M.}~\bibnamefont{Fiebig}},
  \bibinfo{author}{\bibfnamefont{V.~V.} \bibnamefont{Pavlov}},
  \bibnamefont{and} \bibinfo{author}{\bibfnamefont{R.~V.}
  \bibnamefont{Pisarev}}, \bibinfo{journal}{J.~Opt. Soc. Am.~B}
  \textbf{\bibinfo{volume}{22}}, \bibinfo{pages}{96} (\bibinfo{year}{2005}).

\bibitem[{\citenamefont{Degenhardt et~al.}(2001)\citenamefont{Degenhardt,
  Fiebig, Fr{\"o}hlich, Lottermoser, and Pisarev}}]{Degenhardt_et_al:2001}
\bibinfo{author}{\bibfnamefont{C.}~\bibnamefont{Degenhardt}},
  \bibinfo{author}{\bibfnamefont{M.}~\bibnamefont{Fiebig}},
  \bibinfo{author}{\bibfnamefont{D.}~\bibnamefont{Fr{\"o}hlich}},
  \bibinfo{author}{\bibfnamefont{T.}~\bibnamefont{Lottermoser}},
  \bibnamefont{and} \bibinfo{author}{\bibfnamefont{R.~V.}
  \bibnamefont{Pisarev}}, \bibinfo{journal}{Appl. Phys. B}
  \textbf{\bibinfo{volume}{73}}, \bibinfo{pages}{139} (\bibinfo{year}{2001}).

\bibitem[{\citenamefont{Fox and Scott}(1977)}]{Fox/Scott:1977}
\bibinfo{author}{\bibfnamefont{D.~L.} \bibnamefont{Fox}} \bibnamefont{and}
  \bibinfo{author}{\bibfnamefont{J.~F.} \bibnamefont{Scott}},
  \bibinfo{journal}{J.~Phys C} \textbf{\bibinfo{volume}{10}},
  \bibinfo{pages}{L329} (\bibinfo{year}{1977}).

\bibitem[{\citenamefont{Lawes et~al.}(2005)\citenamefont{Lawes, Harris, Kimura,
  Rogado, Cava, Aharony, Entin-Wohlmann, Yildirim, Kenzelmann, Broholm
  et~al.}}]{Lawes_et_al:2005}
\bibinfo{author}{\bibfnamefont{G.}~\bibnamefont{Lawes}}
  \bibnamefont{et~al.}, \bibinfo{journal}{Phys. Rev. Lett.}
  \textbf{\bibinfo{volume}{95}}, \bibinfo{pages}{087205}
  (\bibinfo{year}{2005}).

\bibitem[{\citenamefont{Bradley and Cracknell}(1972)}]{Bradley/Cracknell:Book}
\bibinfo{author}{\bibfnamefont{C.~J.} \bibnamefont{Bradley}} \bibnamefont{and}
  \bibinfo{author}{\bibfnamefont{A.~P.} \bibnamefont{Cracknell}},
  \emph{\bibinfo{title}{The mathematical theory of symmetry in solids}}
  (\bibinfo{publisher}{Oxford University Press}, \bibinfo{year}{1972}).

\end{thebibliography}

\end{document}